\documentclass[showpacs, twocolumn]{revtex4-1}
\usepackage{graphicx}
\usepackage{amsmath}
\usepackage{appendix}
\usepackage{bbold}
\usepackage{csquotes}

\begin{document}

\title{ Quench dynamics of disordered quadrupolar Bose-Einstein condensates}

\author{Abdel\^{a}ali Boudjem\^{a}a$^{1,2}$}
\affiliation{$^1$ Department of Physics, Faculty of Exact Sciences and Informatics, Hassiba Benbouali University of Chlef, P.O. Box 78, 02000, Ouled-Fares, Chlef, Algeria. \\
$^2$Laboratory of Mechanics and Energy, Hassiba Benbouali University of Chlef, P.O. Box 78, 02000, Ouled-Fares, Chlef, Algeria.}
\email {a.boudjemaa@univ-chlef.dz}

\date{\today}

\begin{abstract}

We systematically investigate the equilibrium and the nonequilibrium quench dynamics of three-dimensional disordered quadrupolar Bose-Einstein condensates.
Within the Bogoliubov-Huang-Meng approximation, we show that the combined effect of quenched interactions, disorder and excitations may 
modify the static as well as the dynamic properties of the system.
The validity criterion of the developed approach is accurately established.
By quenching the interaction strength,  we reveal that the quantum depletion and the deformation condensate induced by disorder 
are enhanced in the asymptotic steady state compared to the corresponding equilibrium values.
The time evolution of the condensate deformation is accompanied by damped oscillations with amplitudes strongly depend on 
the disorder correlation length and on the relative  quadrupolar interaction.

\end{abstract}

\maketitle

\section{Introduction} \label{Intro}

In recent years, ultracold quantum gases with either electric or magnetic quadrupole-quadrupole interaction (QQI) have attracted a great theoretical interest
due to their potential applications \cite{Bhon, Lahz1, Malom, Pikov, Lahz}.
Experimentally,  quadrupolar Bose-Einstein condensates (BECs) can readily be created  with alkaline-earth and rare-earth atoms such as Cs$_2$ \cite{Web} and  Sr$_2$ \cite{Stel,Rei}.
The realization of Dy atoms possessing an electric, in addition to the magnetic, dipole moment has been also suggested in \cite{Lepers}.

The peculiar anisotropy and the broad tunability of QQIs make them unique for exploring many interesting aspects in quadrupolar  ultracold  gases including 
quantum phases of quadrupolar Fermi gases in an optical lattice \cite{Bhon}, roton-like excitations \cite{ Lahz}, the ground-state properties \cite{Wang}.
The formation of solitons and quantum droplets in QQI-coupled BEC was addressed in (see e.g. \cite{Malom, Mish,Malom2}). 
However,  to the best of our knowledge, no attempts have been directed toward quadrupolar BECs confined in random media. 
Weakly interacting Bose gases in a random external potential have become recently the subject of many theoretical and experimental investigations  
due to a high degree of control (see for review \cite{Asp,Modu}).
For a dipolar BEC in random potentials, it turns out that the condensed density reduces and the superfluid density acquires 
a characteristic direction dependence \cite{Axel1,Axel2,Axel3, Boudj1,Boudj2,Boudj3,Boudj4}.
In the 2D case, we have shown that the interplay of disorder and rotonization induced by the dipole-dipole interaction (DDI) may strongly reduce the superfluidity leading to a
transition to a superglass phase \cite{Boudj5,Boudj6,Boudj7}.
Furthermore, Bogoliubov excitations of weakly  interacting nondipolar BECs in the presence of disorder have been studied in \cite{Lug, Gaul}.

The aim of this paper is to investigate  the equilibrium and nonequilibrium features of three-dimensional (3D) homogeneous disordered quadrupolar BECs 
following a sudden change in the interaction strength.
Quadrupolar particles moving in a disordered environment may open fascinating prospects for the observation of non-trivial quantum phases, 
because it connects three major ideas namely: disorder, excitations and quenched QQI.

We first study the ground-state properties of quadrupolar BEC subjected to a 3D weak speckle disorder.
We derive useful formulas  for the quantum depletion, the condensate deformation and the equation of state (EoS)  using the Bogoliubov-Huang-Meng approximation (BHMA) \cite{HM}.
This latter has been successfully employed in the analysis of disordered dipolar \cite{Axel3,Boudj1,Boudj2,Boudj3,Boudj4} and nondipolar \cite{HM,Gior,Lug, Gaul} BECs as well as in
the exploration of quantum liquid droplets in random media \cite{Abbas2, Abbas3, Abbas4}. 
Within the BHMA we find that the complex interplay of the disorder and the anisotropy of the QQI may enhance these quantities in the equilibrium regime.
Importantly, we find that at higher temperaures, the thermal fluctuations acquire an interaction dependence which is in stark contrast with the 
dipolar and nondipolar BECs. 
In passing, we recover the properties of disordered dipolar and nondipolar BECs as described in recent papers \cite{Axel3, Boudj1} and in the early HM work \cite{HM}. 

In the second part, we extend the discussion to the out-of-equilibrium behavior arising from a sudden quench of the interactions. 
Recently, quench dynamics of dilute Bose gases in disordered environments has become one of the interesting research subjects since it leads to novel quantum phase transitons
and to the formation of fascinating steady states (see e.g.\cite {ZZ,Scoq,Cher,Rad, Abbas,Nag,Boudj9}).
Here we consider two interaction quench protocols: from noninteracting BEC to an interacting quadrupolar Bose gas and quench between arbitrary  two scattering lengths.
In the former the condensate evolves with a usual Bogoliubov excitations spectum  while in the latter senario, the spectum is modified.
By applying the time-dependent Bogoliubov-Huang-Meng approximation (TDBHMA),  the time-dependent quantum depletion and the  condensate deformation induced by disorder 
are determined as a function of the disorder parameters (i.e. the disorder strength and the correlation length), and the characteristic relaxation time following the quench.
On longer time scales, we show analytically that such a system supports nonequilibrium a steady state for the depletion and the condensate deformation
larger than their corresponding equilibrium values.
We then carry out a detailed analysis of the full time dependence of the quantum depletion and the condensate deformation by numerically solving the obtained TDBHM equations.
It is shown that the quantum depletion increases linearly at short times then it saturates at later times signaling the existence of a prethermalization regime.
The condensate deformation experiences damped radius oscillation, and reduces with QQI preventing the formation of an insulating phase.

The rest of the paper is organized as follows. 
Section~\ref{EP} addresses the BHMA for quadrupolar Bose gases subjected to a speckle potential in the equilibrium regime. 
We derive analytical expressions for the quantum depletion, the condensate deformation and the EoS. Our results are compared with those obtained in the literature.
In Sec.~\ref{NED}  we perform a detailed investigation of the nonequilibrium dynamics of a homogeneous BEC with QQI 
by suddenly quenching the relative interaction strength employing the TDBHMA.
We examine the short- and long-time behaviors of the quantum depletion and the condensate deformation at zero temperature. 
Our reults are summarized in section \ref{conc}.

\section{Equilibrium properties} \label{EP}

\subsection{Bogoliubov-Huang-Meng approximation} \label{BHMT}

We consider a weakly interacting quadrupolar BEC  at zero temperature, subjected to a weak disorder potential $U$. 
The disorder potential is assumed to have vanishing ensemble averages $\langle U({\bf r})\rangle=0$
and a finite correlation of the form $\langle U({\bf r}) U({\bf r'})\rangle=R ({\bf r}-{\bf r'})$. 
The interaction potential is given by:
$V(\mathbf{r}-\mathbf{r^\prime})=g\delta(\mathbf{r}-\mathbf{r^\prime})+V_{\text{qq}}(\mathbf{r}-\mathbf{r^\prime})$, where $g=4\pi \hbar^2 a/m$
corresponds to the short-range part of the interaction with $a$ being the scattering length, and the magnetic QQI 
between two particles having a quadrupole moment $q$ separated by the distance vector $r$ and aligned via an external magnetic field $B$ which is assumed to be
fixed along the $z$-axis, reads 
\begin{align}\label{qqi}
V_{\text{qq}}(\mathbf r)&= \frac{3q^2}{16 \pi \epsilon_0} \frac{3- 30\cos^2 \theta+35 \cos^4 \theta}{r^5} \\
&=g_q \frac{ Y_4^0 (\theta,0)}{r^5}, \nonumber
\end{align}
where $g_q=q^2/ (\sqrt{\pi} \epsilon_0)$ characterizes the strength of the QQI, $\theta$ is the polar angle, and $Y_ 4^0 (\theta)$ is the spherical harmonic.
The QQI changes its character from repulsive for $\theta \in [ 0, \theta_1 =\arccos (15+2 \sqrt{30})/35]$ and $\theta \in [\theta_2 =\arccos (15-2 \sqrt{30})/35, \pi/2]$ 
to attractive for $\theta \in [\theta_1 ,\theta_2]$ \cite{Lahz}.

The Hamiltonian of the system can be written in terms of the creation and annihilation operators $\hat a^\dagger_{\bf k}$ and $\hat a_{\bf k}$ as:
\begin{align}\label{ham1}
\hat H\!\!&=\!\!\sum_{\bf k}\! E_k\hat a^\dagger_{\bf k}\hat a_{\bf k}\! 
+\!\frac{1}{ {\cal V}}\!\!\sum_{\bf k,\bf p} \! U ({\bf k\!-\!\bf p}) \hat a^\dagger_{\bf k} \hat a_{\bf p}  \\
&+\!\frac{1}{2 {\cal V}}\!\!\sum_{\bf k,\bf p,\bf q}\!\, V ({\bf q})\, \!\hat a^\dagger_{\bf k\!+\!\bf q} \hat a^\dagger_{\bf p\!-\!\bf q}\hat a_{\bf p}\hat a_{\bf k}. \nonumber
\end{align}
where $E_k=\hbar^2k^2/2m$, ${\cal V}$ is a quantization volume, $U({\bf k})$ is the Fourier transform of the external random potential $U({\bf r})$, 
and $V(\mathbf k)$ is the Fourier transform of the interaction potential which is written as \cite{Lahz}: 
\begin{equation}\label{intP}
V(\mathbf k)= g \bigg[ 1+\frac{4 \pi}{105} \epsilon_{\text{qq}}  Y_4^0 (\theta) k^2 \bigg],
\end{equation}
where $\theta$ is the azimuthal angle between the vector ${\bf k}$ and the polarization direction, 
and $\epsilon_{\text{qq}}= g_q/g$ denotes the relative interaction strength.  

We assume that almost all molecules are in the ground-state of a dilute Bose quadrupolar BEC, we then replace the operators $\hat a_{0}$ and $\hat a^\dagger_{0}$  
by a $c$-number, i.e., $\hat a_{0}= \hat a^\dagger_{0}=\sqrt{N}$, where $N$ is the number of particles. 
Using the normalization relation: $N= \hat a^\dagger_{0}\hat a_{0}+\sum_{\mathbf k\neq 0}\hat a^\dagger_{\bf k}\hat a_{\bf k}$, the Hamiltonian (\ref{ham1}), takes the form
\begin{align}\label{ham2}
&\hat  H =   \frac{1} {2} N  n V (0)+n U_0 
+\!\frac{\sqrt{ N} }{ {\cal V}} \,\!\sum_{\mathbf k \neq 0} \, U_{\bf k\!}\, \left( \hat a^\dagger_{\bf k} +\hat a_{-\bf k}\right)  \\
&+ \,\!\sum_{\mathbf k \neq 0}\!  \left [ E_k+ V ({\bf k}) n \right]\hat a^\dagger_{\bf k}\hat a_{\bf k}\!  
+ \frac{1}{2} n\,\sum_{\mathbf k \neq 0} V ({\bf k}) \left(\hat a^\dagger_{\bf k} \hat a^\dagger_{-\bf k}+ \hat a_{\bf k} \hat a_{-\bf k} \right), \nonumber
\end{align}
where $n=N/ {\cal V}$ is the density of the gas.
In Eq.~(\ref{ham2}) we kept only quadratic terms in $\hat a^\dagger_{{\bf k} \neq0}$,  $\hat a_{{\bf k} \neq0} $ up to the second-order in the coupling constants.
We also assumed that for weak enough disorder,  disorder fluctuations decouple in the lowest order \cite{HM,Gior}. 
As a result  we ignored the terms $U_{\bf k\!-\!\bf p} \hat a^\dagger_{\bf k} \hat a_{\bf p}$ with both $\mathbf k=0$ and $ \mathbf p=0$.
Hamiltonian (\ref{ham2}) can be diagonalized using the canonical BHM transformation  \cite{HM}
 $\hat a_{k}= (u_{k} \hat b_{k}- v_{k} \hat b_{-k}^\dagger-\beta_{\bf k})$, where 
the quasi-particle annihilation, $\hat b_{\mathbf{k}} $, and creation, $\hat b_{\mathbf{k}}^{\dagger} $, 
operators obey the usual Bose commutation relations: $\left[\hat b_{\mathbf{k}}, \hat b_{\mathbf{k'}}^{\dagger}\right]=\delta_{\mathbf{k,k'}}$,
$\left[\hat b_{\mathbf{k}}, \hat b_{\mathbf{k'}}\right]= \left[\hat b_{ \mathbf{k}}^{\dagger}, \hat b_{\mathbf{k'}}^{\dagger}\right]=0$.
The Bogoliubov quasiparticle amplitudes are defined as:
\begin{equation}\label{Eq12}
 u_{ k}=\frac{1}{2}\left(\sqrt{\frac{\varepsilon_{ k}}{E_k}}+\sqrt{\frac{E_k}{\varepsilon_{ k}}}\right),\;\;\; v_{k}=u_{ k}-\sqrt{\frac{E_k}{\varepsilon_{ k}}},
\end{equation}
which are chosen to be real without loss of generality,  and the disorder translation $\beta_{k}$ is defined by the equation
\begin{align}  \label{Beta}
\beta_{\bf k}=\sqrt{\frac{n}{\cal V}}  \frac{|u_{k}-v_{k}|^2} {\varepsilon_{k}} U_{\bf k}.
\end{align}
The above BHM transformation allows us to decouple the quantum and random variables and get the bilinear form of the Hamiltonian, which can be written as:
$\hat  H\!\!= E+ \sum\limits_{{\bf k}\neq0} \varepsilon_{k} \hat b^\dagger_{\bf k} \hat b_{\bf k}$, 
where 
\begin{equation}\label{Bogspec}
\varepsilon_{k}= \sqrt{E_k^2 + 2E_k n g \left[1+ \frac{4 \pi \epsilon_{\text{qq}}}{105} Y_4^0 (\theta_{\bf k}) k^2 \right]},
\end{equation}
is the Bogoliubov spectrum.
For small momenta, $\varepsilon_{k}= \hbar c_s k$, with $c_s=\sqrt{gn/m} $ being the sound velocity. 
Remarkably, the sound velocity is isotropic (independent on the direction of the vector $\mathbf k$) in stark contrast to the standard dipolar BEC.
Therefore, a homogeneous quadrupolar BEC suffers from phonon instability if $g<0$.
In the high momenta limit, one has $\varepsilon_{k}= \sqrt{(\hbar^2 /2m)^2 + (4 \pi \hbar^2 n g_q  /105 m) Y_4^0 (\theta_{\bf k})} k^2$,
so the condition for stability of the system requires that the kinetic energy to be larger than or equal to any attractive QQI, i.e.
$(\hbar^2 /2m)^2 + (4 \pi \hbar^2 n g_q  /105 m) Y_4^0 (\theta_{\bf k}) k^2>0$. 

In the thermodynamic limit where the sum over $k$ can be replaced by the integral $\sum_{\bf k}={\cal V}\int d^3k/(2\pi)^3$, the ground-state energy takes the form:
\begin{equation}\label{genergy} 
E= E_0+ E_{\text{LHY}}+ E_R,
\end{equation}
where $E_0/{\cal V}=n^2  g/2$ is the mean-field energy,
\begin{equation}\label{engyLHY} 
\frac{E_{\text{LHY}}}{\cal V}=\frac{1}{2} \int \frac{d^3 k}{(2\pi)^3}\, \left(\varepsilon_{k} -2E_k- n g \right), 
\end{equation}
accounts for the Lee-Huang-Yang (LHY) quantum corrections to  the ground-state energy, and
\begin{equation}\label{Rengy} 
\frac{E_R}{\cal V}=- n \int \frac{d^3 k}{(2\pi)^3} \  R_{\bf k} \frac{ E_k}{\varepsilon_{k}^2},
\end{equation}
denotes the disorder corrections to the ground-state energy, where $R_{\mathbf k}$ is the
Fourier transform of the disorder correlation function $R(\mathbf r)$.

The noncondensed density is defined as  $\tilde{n}={\cal V}^{-1}\sum_{\bf k} \langle\hat a^\dagger_{\bf k}\hat a_{\bf k}\rangle$. 
Then invoking for the operators $\hat a_{\bf k}$ the above Bogoliubov transformation, and using $\langle \hat b^\dagger_{\bf k}\hat b_{\bf k}\rangle=N_{\bf k}$,
where $N_{\bf k}=[\exp(\varepsilon_k/T)-1]^{-1}$ are occupation numbers for the excitations, 
and $\langle \hat b^\dagger_{\bf k}\hat b^\dagger_{\bf k}\rangle=\langle \hat b_{\bf k}\hat b_{\bf k}\rangle=0$, we obtain:
\begin{equation}\label {norr}
\tilde{n}= \tilde{n}_0+\tilde{n}_T+n_R,
\end{equation}
where
\begin{equation}\label {nor}
\tilde{n}_0=\frac{1}{2}\int \frac{d^3k} {(2\pi)^3} \left[\frac{E_k+V({\bf k}) n} {\varepsilon_k}-1\right],
\end{equation}
denotes the condensate depletion due to interaction and quantum fluctuations, and
\begin{equation}\label {norT}
\tilde{n}_T=\frac{1}{2}\int \frac{d^3k} {(2\pi)^3} \frac{E_k+V({\bf k}) n} {\varepsilon_k} N_k,
\end{equation}
accounts for  the condensate depletion due to thermal fluctuations.
The last term in Eq.~(\ref{norr}) represents the density of the disorder-averaged condensate (or the condensate deformation)
(i.e. condensate depletion induced by the disorder potential) which is defined as \cite{HM}      
\begin{equation} \label{glass}
n_R={ \cal V}^{-1}\sum_{\mathbf k \neq 0} \langle |\beta_{\bf k}|^2 \rangle= n \int_0^{\infty}  \frac{ d^3k}{ (2\pi)^3} \frac{ E_k^2}{ \varepsilon_{k}^4} R_{\bf k}.
\end{equation}
It is obtained by performing the disorder ensemble average.

\subsection{Results} \label{res}

Let us now consider a quadrupolar BEC subjected to a 3D random potential generated by a laser speckle that is commonly used with ultracold atoms 
(see e.g. \cite{Asp,Modu,Bily, Roat, Chen}).
Experimentally, an isotropic 3D speckle, can be achieved as the interference pattern of many wavevectors inside a closed optical cavity \cite{Asp,Modu}.
The speckle disorder is characterized by the following real-space correlation function : 
$R (\mathbf r)=R\, \mathrm{sinc}^2(\mathbf r /\sigma)$, where $R=U_R^2$ is the disorder strength and $\sigma$ is the disorder correlation length.
The corresponding Fourier transform has the form \cite{Kuhn}:
\begin{equation}  \label{speckle}
R_{\mathbf k}= \frac{\pi ^2 R}{ k \sigma} \Theta\left(2- k\sigma\right),
\end{equation}
where $\Theta $ denotes the Heaviside function. 

\begin{figure}[h]
\centerline{
\includegraphics[scale=0.46]{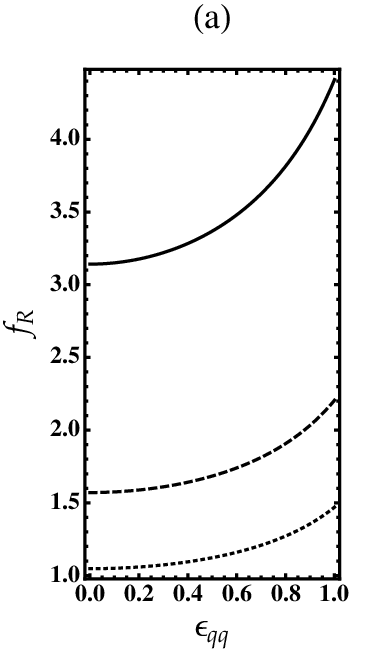}
\includegraphics[scale=0.46]{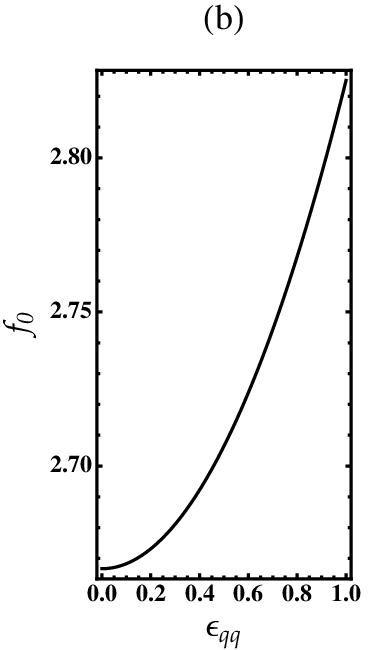}
\includegraphics[scale=0.46]{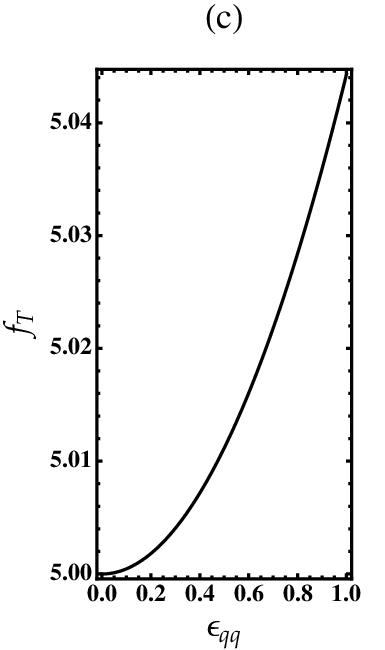}}
 \caption{ Behavior of the functions (a) $f_R$ for different values of $\sigma/\xi$ (solid line: $\sigma/\xi=0.5$, dashed line: $\sigma/\xi=1$, dotted line: $\sigma/\xi=1.5$), 
(b) $f_0$, and (c) $f_T$ as a function of the relative interaction strength, $\epsilon_{\text{qq}}$.}
\label{GL} 
\end{figure}

The condensate deformation due to disorder effects can be computed via Eq.~(\ref{glass})
\begin{align}\label {norR}
n_{R} = n_{\text{HM}} f_R(\sigma/\xi,\beta(\theta)),
\end{align}
where $n_{\text{HM}}= R \, m^2 \sqrt{n/a}/ \left(8\pi^{3/2} \hbar^4\right)$ is the standard HM result for the isotropic contact interaction \cite{HM}
which was obtained through an adiabatic switching on of the disorder, and 
$$f_R(\sigma/\xi,\beta(\theta))= \frac{\pi}{2} \frac{\xi } {\sigma} \int_0^\pi \frac{\sin \theta d \theta}  {  1+ 4 \beta (\theta) },$$
with $\xi =\hbar/\sqrt{mng}$ being the healing length and $\beta(\theta)=1/4+  4 \pi \epsilon_{\text{qq}} Y_4^0 (\theta) /(105 \xi^2)$.
The behavior of the function $f_R$ is depicted in Fig.~\ref{GL} (a).
We see that $f_R$ increases monotonically with both $\epsilon_{\text{qq}}$ and $(\sigma/\xi)^{-1}$ which may lead to enhance the fraction of the condensate deformation and thus, 
reduce the condensed fraction.

The quantum depletion of Eq.~(\ref{nor}) takes the form:
\begin{equation}\label {nor1}
\tilde{n}_0= \frac{1}{8 \pi^2 \xi^3} f_0( \beta(\theta)),
\end{equation}
where 
$$f_0( \beta(\theta))=\int_0^\pi \sin \theta d \theta \bigg[4 \sqrt{\beta(\theta)} -\frac{2 \beta(\theta) -1}{6 \beta^2(\theta)}+\frac{1}{\sqrt{\beta(\theta) }}-4 \bigg],$$
its behavior is displayed in Fig.~\ref{GL} (b). We observe that $f_0$ is rising with $\epsilon_{\text{qq}}$ signaling that the quantum depletion of quadrupolar BEC becomes important.
For $\beta(\theta)= 1/4$, one has $f_0=8/3$ hence, the quantum depletion reduces to that of a dilute BEC with contact interactions namely: $\tilde{n}_0= 1/(3\pi^2 \xi^3)$.

At finite temperatures, the thermal depletion (\ref{norT}) will play a crucial role. 
At very low temperatures, $\mu \ll ng$, the main contribution to the integral (\ref{norT}) comes from the excitations of the phonon branch, therefore, 
the thermal depletion takes its standard form: $\tilde{n}_T= m T^2/(12 \hbar^3 c_s)$.
However at temperatures $\mu \gg ng$, where the main contribution to the integral (\ref{norT}) comes from single-particle excitations, the thermal depletion reads
 \begin{equation}\label {norT1}
\tilde{n}_T \approx  \frac{\zeta(3/2)}{2} \left(\frac{mT}{2\pi \hbar^2}\right)^{3/2} f_T( \beta(\theta)),
\end{equation}
where $\zeta(3/2)$ is the Riemann zeta function, and 
$$ f_T( \beta(\theta))= \int_0^\pi \sin \theta d \theta \bigg[  \frac{4\beta^2(\theta)+1 }{2\beta(\theta)}\bigg].$$
This extra term represents the QQI contribution to the thermal fluctuations.
Surprisingly,  Eq.~(\ref{norT1}) shows that the thermal fluctuations of a quadrupolar BEC depend on the interactions.
Such a dependence on the interaction is fully absent in the usual dipolar and nondipolar Bose gases at higher temperatures \cite{Axel1,Axel2,Axel3, Boudj1,Boudj2,Boudj3,Boudj4,Boudj8}.
For $\beta(\theta)= 1/4$, $\tilde{n}_T$ reduces to the density of an ideal Bose gas.
As is seen in Fig.~\ref{GL} (c), the presence of QQIs may strongly enhance the thermal fluctuations.

The BHMA assumes that fluctuations arsing from both interaction and disorder must be small. We then deduce from Eqs.~(\ref{norR}) and  (\ref{nor1})  that at $T = 0$, 
the validity of the BHMA requires inequalities $ f_0( \beta(\theta))/(8 \pi^2 \xi^3) \ll 1$ and $ n_{\text{HM}} f_R(\sigma/\xi,\beta(\theta)) \ll 1$, or equivalently 
$R< R_c= 8\pi^{3/2} \hbar^4/ \left[m^2 \sqrt{n/a} f_R(\sigma/\xi,\beta(\theta)) \right] $, where $R_c$ is the critical disorder strength beyond which  the BHMA is not applicable.
For $R=0$, the above parameter differs only by the factor $f_0( \beta(\theta))$ from the universal small parameter of the theory, $\sqrt{na^3}\ll 1$, in the absence of QQI. 
At low temperatures, the Bogoliubov theory requires an extra condtion $(T/gn) \sqrt{n a^3} \ll 1$, which comes from the thermal fluctuations corrections.
However at higher temperatures the BHMA is no longer valid.

\begin{figure}[h]
\includegraphics[scale=0.65]{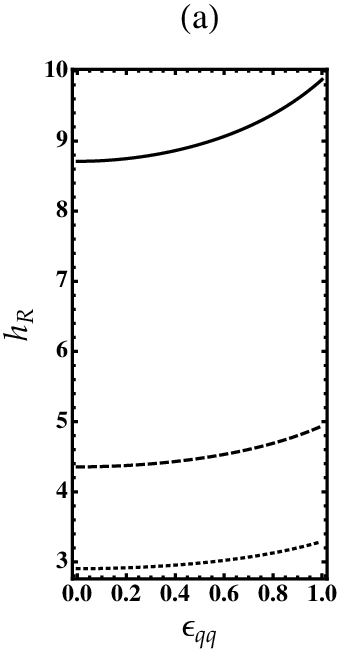}
\includegraphics[scale=0.65]{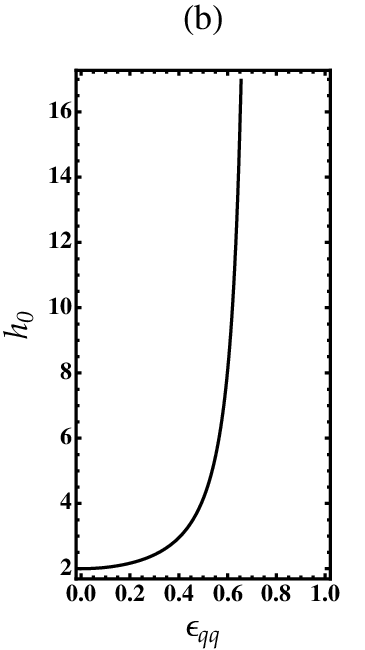}
 \caption{ Behavior of the functions (a) $h_R$ for different values of $\sigma/\xi$ (solid line: $\sigma/\xi=0.5$, dashed line: $\sigma/\xi=1$, dotted line: $\sigma/\xi=1.5$), 
and (b) $h_0$ as a function of the relative interaction strength, $\epsilon_{\text{qq}}$.}
\label{ER} 
\end{figure}

The energy shift due to disorder corrections  can be computed through (\ref{Rengy}). This yields
\begin{align}\label{Rengy1} 
\frac{E_R}{N}=-  \frac{R m \sqrt{na/\pi}}{\hbar^2}  h_R( \beta(\theta)),
\end{align}
where 
$$h_R( \beta(\theta))= \frac{\pi}{4}\frac{\xi } {\sigma} \int_0^\pi \sin \theta d \theta \bigg[  \frac{ \ln (1+4\beta(\theta)) }{2\beta(\theta)}\bigg].$$
In Fig.~\ref{ER} (a) we plot the function $h_R( \beta(\theta))$ as a function of the relative interaction strength, $\epsilon_{\text{qq}}$.
Here we see also that $E_R/N$ increases with both $\epsilon_{\text{qq}}$ for any $\sigma/\xi$.
The negative sign of $E_R/E_0$ indicates that the disorder contribution leads to a decrease in the ground-state energy.
The most likely cause for this is an additional scattering induced by the random potential.

By integrating the expression (\ref{engyLHY}) over momentum we obtain for the LHY-corrected EoS
\begin{equation}\label{engyLHY1} 
\frac{E_{\text{LHY}}}{E_0}=\frac{1}{4 \pi^2 \xi^3 n} h_0( \beta(\theta)), 
\end{equation}
where
$$h_0( \beta(\theta))= \int_0^\pi \sin \theta d \theta \bigg[ \frac{2}{15 \beta^2(\theta)}+2 \sqrt{\beta(\theta)}-1\bigg],$$
is the QQI correction to the energy.
It is noteworthy that in the absence of the disorder, $R=0$, and for $\beta(\theta)=1/4$, one has $h_0=2$ (see Fig.~\ref{ER} (b)), thus the ground-state energy 
simplifies to the usual LHY form: $E/E_0= 1+ (128 /15) \sqrt{na^3/\pi}$ \cite{LHY}.
Figure \ref{ER} (b) shows also that the QQIs lift the LHY-corrected energy.

\section{Non-equilibrium dynamics} \label{NED}

\subsection{Time-dependent Bogoliubov-Huang-Meng approximation} \label{tdBHMT}

We now use the TDBHMA for studying the dynamics induced by the interaction quench of  weakly interacting disordered quadrupolar BEC  at zero temperature.
The TDBHMA has been widely employed to describe the dynamic quench for both clean and disordered BECs  (see e.g. \cite{ZZ,Natu}).
It is a generalization of the above BHMA, where the canonical transformation becomes time-dependent
\begin{equation}\label {trans}
 \hat a_{\bf k} (t)= u_k(t) \hat b_{\bf k}-v_k(t) \hat b^\dagger_{-\bf k}-\beta_{\bf k} (t), 
\end{equation}
where
\begin{align}  \label{BetaTD}
\beta_{\bf k} (t)=\sqrt{\frac{n}{\cal V}}  \frac{|u_{k} (t)-v_{k} (t)|^2} {\varepsilon_{k}^f} U_{\bf k},
\end{align}
and $u_k(t)$ and $v_k (t)$ are time-dependent Bogoliubov amplitudes acquiring complex values and satisfying the standard normalization condition: $|u_k(t)|^2-|v_k(t)|^2=1$.
Their equations of motion are obtained from the Heisenberg equations of motion for $a_{\mathbf k}$ \cite{Natu,ZZ}. This yields:
\begin{align} \label{BdGE2}
		\begin{pmatrix} 
		u_{k} (t)\\
		v_{k}(t)\\
		\end{pmatrix}
&= \bigg[\cos (\varepsilon_k^f t) \mathbb{1} -i \frac{\sin (\varepsilon_k^f t) } {\varepsilon_k{^2}^f} \\
&\times\begin{pmatrix}
		E_k+ V^f(\mathbf k) n & V^f (\mathbf k) n&\\
		-V^f (\mathbf k)  n &- E_k- V^f (\mathbf k) n)&\\
		\end{pmatrix}
		\bigg]
		\begin{pmatrix}
		u_{k} (0)\\
		v_{k}(0)\\
		\end{pmatrix} ,
		\nonumber
\end{align} 
where $\mathbb{1}$ is the identity matrix, 
\begin{equation}\label{intF}
V^f(\mathbf k)= g^f \bigg[ 1+\frac{4 \pi}{105} \epsilon_{\text{qq}}^f  Y_4^0 (\theta) k^2 \bigg],
\end{equation}
with $\epsilon_{\text{qq}}^f= g_q^f/g^f$, and
\begin{equation}\label {TDspec}
\varepsilon_k^f=\sqrt{E_k^2+ 2 V^f(\mathbf k) E_k},
\end{equation}
is the final Bogoliubov excitations spectrum.  

The time-dependent noncondensed  density is then given by $\tilde{n} (t)= \tilde{n}_0 (t)+n_R (t)$, 
where $ \tilde{n}_0 (t)={ \cal V}^{-1}\sum\limits_{\mathbf k \neq 0} |v_k(t)|^2$ and $n_R(t)={ \cal V}^{-1}\sum\limits_{\mathbf k \neq 0} \langle |\beta_{\bf k}|^2 (t)\rangle$.
By substituting Eqs.~(\ref{BetaTD}) and (\ref{BdGE2}) into the definition of $ \tilde{n}_0 (t)$ and $ n_R (t)$, we arrive at
\begin{equation}\label {TDnor}
\tilde{n}_0 (t)=  \tilde{n}_0+ \int \frac{d^3k}{(2\pi)^3} V^f(\mathbf k)[V^f(\mathbf k)- V(\mathbf k)] n^2 E_k \frac{\sin^2 (\varepsilon_k^f t)}{\varepsilon_k^{2f} \varepsilon_k}.
\end{equation}
and
\begin{align}  \label{TDglass}
n_R (t)&=n_R+4 n \int_0^{\infty} \frac{d^3k}{(2\pi)^3} (V^{f} (\mathbf k) n) \frac{\sin^2(\varepsilon_k^f t) E_k^2 } {\varepsilon_k^{f2} \varepsilon_k^2} \\
& \times\big(E_k+V(\mathbf k) n \big) \bigg[\frac{V^f (\mathbf k) n(E_k+V(\mathbf k) n)} {\varepsilon_k^{f2} \varepsilon_k^2}+\frac{1}{\varepsilon_k^{2}} \bigg] R_{\bf k} \nonumber\\
&-n \int_0^{\infty} \frac{d^3k}{(2\pi)^3} (V^{f} (\mathbf k) n)^2 \frac{\sin^2(2\varepsilon_k^f t) E_k^2 } {\varepsilon_k^{f4} \varepsilon_k^4} \big(E_k+V(\mathbf k) n \big)^2 R_{\bf k}.\nonumber
\end{align}

In what follows we will model the quench of the QQI using two configurations.

\subsection{Quench from $V({\mathbf k})=0$ to $V^f({\mathbf k})>0$}

Let us assume that the system is initially in its equilibrium state i.e., $V({\mathbf k})=0$, when $t\leq 0$, we then suddenly quench its interaction to $V^f({\mathbf k})>0$ when $t> 0$,
and control the out-of-equilibrium dynamics by solving the above TDBHM equations.
The temporal quantities are rescaled via  $t \rightarrow t/\tau$, where $\tau= \hbar /(g^fn)$ denotes the characteristic relaxation time following the quench,
and the momentum is rescaled according to $k \rightarrow k \xi^f$, where $\xi^f= \hbar/\sqrt{mng^f}$ is the healing length after quench.

\begin{figure}
\includegraphics[scale=0.8] {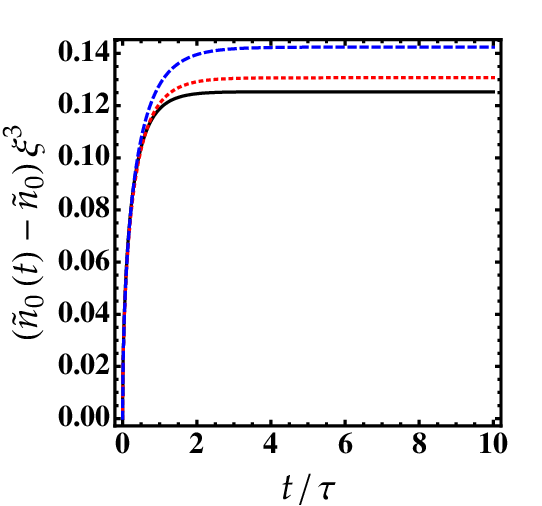}
 \caption{ Time-dependent normalized quantum depletion, $(\tilde n_0(t)-\tilde n_0) \xi^{f3}$, for different values of  the relative interaction strength, $\epsilon_{\text{qq}}^f$.
Black solid lines: $\epsilon_{\text{qq}}^f=0.1$. Red dotted lines: $\epsilon_{\text{qq}}^f=0.5$. Blue dashed lines: $\epsilon_{\text{qq}}^f=1$.}
 \label{disTD1}
\end{figure}

At long times, the main contribution to the above intgrals comes from low momenta regime \cite{Natu}.
Therefore, the quantum depletion reduces to:
\begin{equation}\label {TDnor2}
\tilde{n}_0 (t \rightarrow \infty) \sim  \frac{1}{8\pi \xi^{f3}} S(t),
\end{equation}
where 
$$S (t)= \int_0^{\pi}  \sin \theta  d \theta  \bigg[\frac{(4 \beta (\theta) +3)^2}{64 \sqrt{\beta (\theta)}} \left(1-e^{-\frac{2 t}{ \tau\sqrt{\beta(\theta) }}}\right)\bigg].$$
Obviously, for $t > \tau\sqrt{\beta(\theta)}/2$, Eq.~(\ref{TDnor2}) becomes time-independence and universal 
(i.e. depends only on the small parameters of the theory and on $\epsilon_{\text{qq}}$)
indicating the emergence of a prethermalized state governed by the quadratic many-body Hamiltonian of Eq.~(\ref{ham2}) (see also Fig.~\ref{disTD1}).
Note that such prethermalization phenomenon happens also for ordinary dilute Bose gases with and without disordered potentials out of equilibrium \cite{Scoq,Reg,Duv}.
In the absence of the QQI, (i.e. $\epsilon_{\text{qq}}^f=0$), the function $S$ reduces to $S(t) \sim 2 ( 1- e^{-4t/\tau})$ 
which saturates to a constant value at times $t \gtrsim \tau$. 
This means that the creation of excitations (depleted state) is larger compared to the corresponding equilibrium zero temperature value. 
At very short times, the quantum depletion spreads linearly  as: $\tilde{n}_0 (t) \sim A (8\pi \xi^{f3})  (t/\tau)$, where $A$ is a constant depending on $\epsilon _{\text{qq}}$.  

\begin{figure}
\includegraphics[scale=0.46] {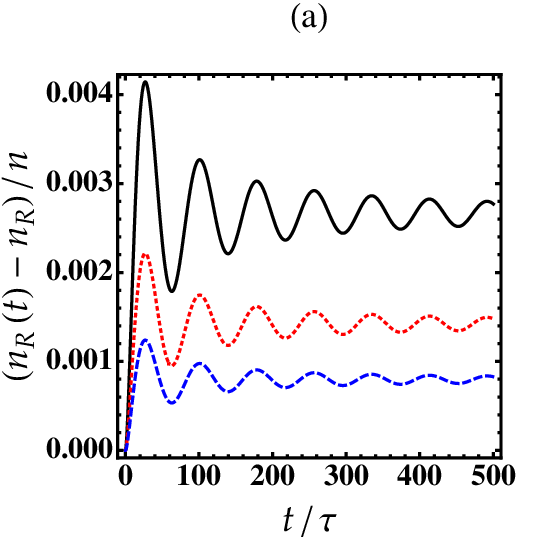}
\includegraphics[scale=0.46] {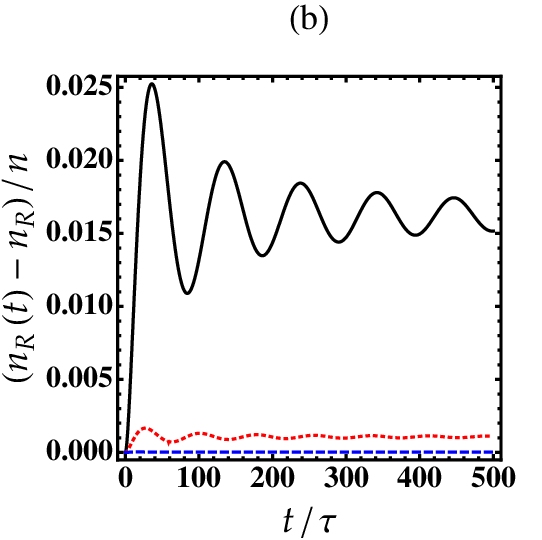}
 \caption{Time-dependent normalized  condensate deformation, $(n_R(t)-n_R)/n$, (a) for different values of  the relative interaction strength, $\epsilon_{\text{qq}}^f$, 
with  $\sigma/\xi^f=1/4$ and $R=003$.
Black solid lines: $\epsilon_{\text{qq}}^f=0.1$, red dotted lines: $\epsilon_{\text{qq}}^f=0.5$, blue dashed lines: $\epsilon_{\text{qq}}^f=1$.
(b) The same but for different values of the disorder correlation length $\sigma/\xi^f$ with $\epsilon_{\text{qq}}^f=0.5$ and $R=003$.
Black solid lines: $\sigma/\xi^f=1/3$, red dotted lines: $\sigma/\xi^f=1/4$, blue dashed lines: $\sigma/\xi^f=1/6$.}
 \label{disTD2}
\end{figure}

In the long-time regime, the condensate  deformation asymptotes to:
\begin{equation}\label {TDglass2}
n_R (t \rightarrow \infty) \sim n_{\text{HM}} S_R(t),
\end{equation}
where the disorder function, after neglecting some unimportant terms, reads
\begin{widetext}
\begin{align}  \label{TDglass}
S_R(t)& \approx \frac{\pi}{8} \int_0^{\pi}  d \theta \sin \theta  \left(\frac{3}{4}+\beta(\theta)\right)^2 
\Bigg \{  \frac{(4 \beta(\theta)  (8 \beta(\theta) +15)+11) \left(\cos \left(\frac{4 t}{\tau }\right)-1\right)}{4 \beta(\theta) +1} +4 e^{-\frac{2 t}{\sqrt{\beta(\theta) } \tau }} 
\bigg[2 \beta (\theta) (4 \beta (\theta)+7) \ln (4 \beta (\theta)+1) \nonumber\\
&-4 \beta (\theta) (4 \beta(\theta) +7) \ln 2+4 \text{Ci}\left(\frac{4 t}{\tau }\right) \left(\beta (\theta) (4 \beta (\theta)+7)+\frac{(4 \beta (\theta)+11) t^2}{\tau ^2}\right)
-\frac{16 (4 \beta (\theta) +11) t^2 \ln \left(\frac{t}{\tau }\right)}{\tau^2}-(4 \beta (\theta)+11)    \nonumber\\
&\times\sin \left(\frac{4 t}{\tau }\right)\bigg]- \text{Ci}\left(\frac{8 t}{\tau }\right) \left(\beta(\theta) +\frac{4 t^2}{\tau ^2}\right)-8 \left[\beta(\theta)  \ln \left(\frac{1}{4} (4 \beta(\theta) +1)\right)
+\frac{8 t^2 \ln \left(\frac{t}{\tau }\right)}{\tau ^2}+\frac{t \sin \left(\frac{8 t}{\tau }\right)}{\tau }\right]\nonumber\\
&+\frac{(8 \beta (\theta)+1) \left(\cos \left(\frac{8 t}{\tau }\right)-1\right)}{4 \beta(\theta) +1} \Bigg\}, \nonumber
\end{align}  
\end{widetext}
where $\text{Ci}(x)$ is the cosine integral function.
Equation (\ref{TDglass2}) reveals that on a time scale $ t=\sqrt{\beta(\theta)}\tau $, the condensate deformation, $n_R(t)$, 
in the asymptotic steady state is enhanced compared to its equilibrium state and exhibits an oscillating behavior (see Fig.~\ref{disTD2}).

To gain deep insights into the combined effect of the disorder and QQI on the quantum depletion and the deformation  condensate, we solve numerically the 
TDBHM equations (\ref{TDnor}) and (\ref{TDglass}) assuming the full Bogoliubov dispersion. The results are displayed in Figs.~\ref{disTD1} and \ref{disTD2}.

Figure~\ref{disTD1} shows that at short times, $t \lesssim \tau$, the normalized quantum depletion is growing linearly with time most probably due to the phonon modes, 
then it saturates to a stationary value at $t \gtrsim 2\tau$ revealing the prethermalization of the molecular gas to a nonthermal steady state.
The  prethermalization time, $t_{\text {preth}}$, increases with QQI, $\epsilon_{\text{qq}}$, for instance $t_{\text {preth}} \simeq 1.8 \tau$ for $\epsilon_{\text{qq}}=0.1$ while
it shifts to $t_{\text {preth}} \simeq 3.2 \tau$ for $\epsilon_{\text{qq}}=1$.
We observe also that $(\tilde n_0(t)-\tilde n_0) \xi^{f3}$ increases with $\epsilon_{\text{qq}}$ notably at longer times which is in fact natural due to the impact of excitations.

Figures \ref{disTD2} (a) and \ref{disTD2} (b) depict that at early times the condensate deformation, $(n_R(t)-n_R)/n$ increases sharply then it experiences damped oscillations 
of period $\sim \pi/(c_s t)$. 
These oscillations which are sizeable at short times, originate from the interference between phonons.
Lowering the disorder correlation length  and increasing the relative srength of QQI might lower the condensate deformation
giving rise to the emergence of a significant fraction of condensed molecules.
One can explain the decay of the condensate deformation as the consequence of a screening of disorder by QQIs.

\subsection{Arbitrary quench}

\begin{figure}
\includegraphics[scale=0.8] {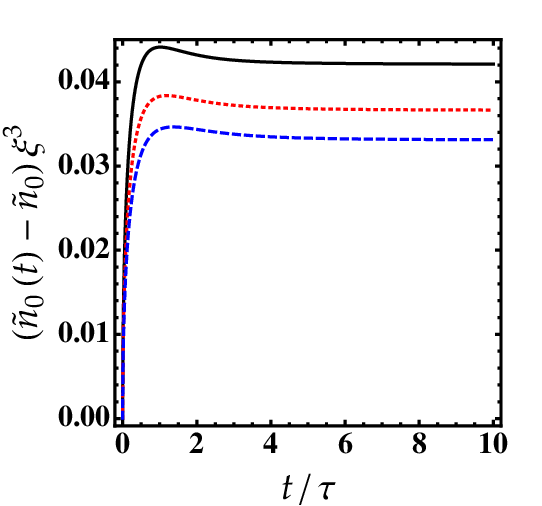}
 \caption{ The same as in Fig.~\ref{disTD1} but for an arbitrary quench.
Parameters are: $\epsilon_{\text{qq}}=0.25$ and $g/g_f=0.5$.}
 \label{disTD3}
\end{figure}

\begin{figure}
\includegraphics[scale=0.46] {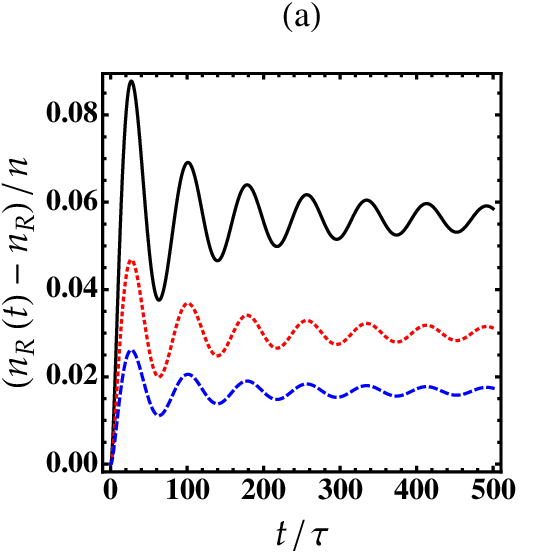}
\includegraphics[scale=0.46] {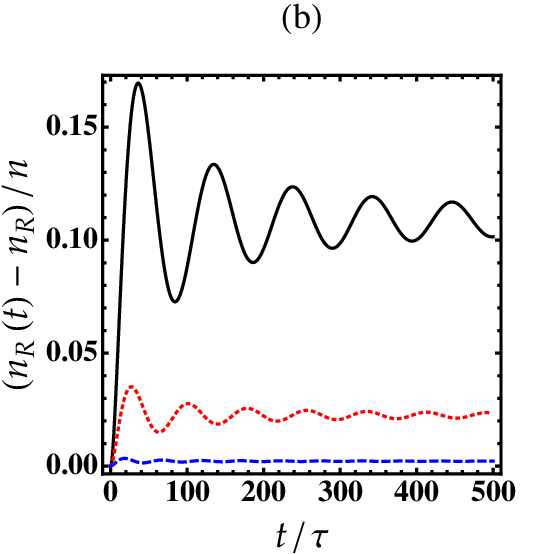}
 \caption{ The same as in Fig.~\ref{disTD2} but for an arbitrary quench.
Parameters are: $R=003$, $\epsilon_{\text{qq}}=0.25$, and $g/g_f=0.5$.}
 \label{disTD4}
\end{figure}

Here, we consider a quench from some initial relative QQI strength, $\epsilon_{\text{qq}}$, to a final value $\epsilon_{\text{qq}}^f$ and 
assume that the $s$-wave scattering length is not affected during the quench.
Experimentally, such an interaction quench can be achieved by using Feshbach resonance.
The full time dependence of the the quantum depletion and the condensate deformation can be obtained via a direct numerical 
simulation of the TDBHM equations (\ref{TDnor}) and (\ref{TDglass}). The results are captured in Figs.~\ref{disTD3} and \ref{disTD4}.

Unlike the previous quench, we can distinguish in Fig.~\ref{disTD3} three stages in the dynamics of the normalized interaction-induced quantum depletion. 
In the first stage, $ t< \tau$, $(\tilde n_0(t)-\tilde n_0) \xi^{f3} \propto t$ which means that in this very short time the depletion expands linearly as can be already seen from Eq.~(\ref{TDnor2}). 
The second stage corresponds to $ t\sim \tau$, where $(\tilde n_0(t)-\tilde n_0) \xi^{f3}$ develops a peak due to the considerable creation of excitations induced by QQI.
Finally, a relaxation to a prethermal plateau is observed in the depletion on a time scale $t  \gtrsim (4\tau)$.
Suprinsingly, the quantum depletion reduces with $\epsilon_{\text{qq}}$  due to the attractive part of the QQI which becomes prominent during the quenched dynamics.

Similarly to the previous case, we observe in Fig.~\ref{disTD4} that the condensate deformation exhibits an oscillatory behavior owing to the interference between phonons.
Amplitude of oscillation, however, displays crucial dependence on QQI and on disorder correlations. 
We see also that $(n_R(t)-n_R)/n$ is larger compared to the previous quench giving rise to a reduction in the fraction of condensed quadrupolar molecules.
Therefore, the quench dynamics may render the disorder capable to deform the condensate more than to deplete it.

\section{Conclusions}\label{conc}

In this paper we presented an in-depth analysis of  equilibrium and nonequilibrium quench dynamics behaviors 
of a homogeneous quadrupolar BEC in a weak 3D speckle disorder, thereby emphasizing the role of quenched interaction, disorder and excitations. 
In the equilibrium regime, we pointed out that the quantum depletion, thermal fluctuations, the condensate deformation, and the EoS are increasing with QQI 
and decaying with the disorder correlation length.
Furthermore, we analyzed  the nonequilibrium dynamics of quantum fluctuations and the condensate deformation in the TDBHMA using two interaction quench scenarios.
In both cases, we showed analytically and numerically that such quantities increase linearly with time in the limit of short times, 
while they reach a prethermal steady state at long times. However an oscillatory behavior is present in the condensate deformation due to the interference between phonons.
One can expect that the anisotropy of QQI would pass on the superfluidity like in the case of disordered dipolar BECs \cite{Axel1,Axel2,Axel3, Boudj1,Boudj2,Boudj3,Boudj4}.
Our results revealed that the delicate interplay of QQI, disorder and quenched excitations helps to control the quantum, thermal and disorder fluctuations.
An interesting question concerns the role of strong time-(in)dependent disorder in the dynamical condensate deformation of quadrupolar BECs.


\subsection*{Data availability statement}
The data generated and/or analyzed during the current study are not publicly available for legal/ethical reasons
but are available from the corresponding author on reasonable request.

\end{document}